# Variable polarisation and luminosity for microlensing of extended stellar sources

J.F.L. Simmons, A.M. Newsam and J.P. Willis
*Department of Physics and Astronomy, University of Glasgow, Glasgow, G12 8QQ, Scotland*

Now

**ABSTRACT**
Microlensing of extended stellar sources in the LMC and Galaxy by low mass lenses can produce variable polarisation. The characteristics of the polarisation and flux profiles can provide considerable information about the lens geometry, and help determine the importance of such low mass objects as a dark matter component.

**Key words:**　gravitational lensing – polarisation – dark matter

## 1　INTRODUCTION

Recently the search for gravitational microlensing events has resulted in a number of successful identifications of lensing events of stars in the LMC and the Galactic Bulge. In the case of the first event reported by the MACHOs group, it would seem that a low mass and possibly compact object in the Galactic Halo was responsible for the microlensing (Alcock et al. 1994). The determination of the mass and spatial density distribution of such objects would have important implications for assessing their relative importance as a dark matter component (Paczynski 1986). The principal effect of lensing is amplification of the flux, which essentially depends on the distance in units of the Einstein radius of the star (source) from the lensing object projected onto the source plane. Various methods have been suggested for determining the Einstein radius of the lens, the most promising being parallax (Gould 1994a), measurement of spectral shifts for rotating stars (Maoz and Gould 1994), and analysis of the light curve (Nemiroff and Wickramasinghe 1994). The latter two effects depend on the star being considered as an extended source. (Even when the radius of the star is considerably less than the Einstein radius these effects can be important). Since the Einstein radius for a given source distance depends on the distance and mass of the lens it is impossible to determine both the lens distance and mass from the light curve. Of course probabilistic, that is maximum likelihood, arguments can be used, although any inferences made in this way will be subject to large uncertainties.

If the source star diameter is larger than a few Einstein radii projected onto the source plane, there will be only a small variation in the flux amplitude during the lensing event. This will make detection of low mass lenses difficult. The fact that one has to consider extended sources could potentially provide more information about the lensing object, and could under certain assumptions such as the star's radius and the transit velocity of the lens, yield both the mass and the distance of the lens (Nemiroff (1994), Witt and Mao (1994), Gould (1994b)). This could be achieved not only by the analysis of the light curve, but more interestingly, by studying the variable polarisation induced by the gravitational lensing (Simmons et al. 1994), and in this paper we address this question.

The probability that any given star is lensed is extremely small. The surveys being carried out depend on monitoring millions of stars, and selecting candidate events on the grounds of a number of criteria, the obvious ones being that the light curve is not periodic, and has the expected profile of a lensing event. Potentially the statistics of the lensing events could yield a lot more information about the spatial and mass distribution of lensing objects, although the interpretation of these statistics will depend on the validity of the assumptions made about the pointlike nature of the source star.

There are three main effects of the star being an extended source that we shall discuss in this paper. The first has been considered recently in relation to MACHOs by Gould (1994a), and by Nemiroff (1994), Witt and Mao (1994).

(i) There will be a modification of the amplification function and thus of the light curve of a lensing event. Our calculation show that for a source star of radius $R > 2.5\eta_0$, where $\eta_0$ is the Einstein radius projected onto the source plane (see fig 3) the amplification is always less than 1.34 (the amplification for a point source at one Einstein radius), no matter how close the lens passes by the centre of the star. This could be crucial for detecting low mass lenses. Indeed it maybe impossible to detect these.
On the other hand for $R < d$, where $d_0$ is the impact parameter on the source plane, the maximum amplification during an event is greater than for a point source. This might explain the light curve anomaly for the first event reported by the MACHOs group, which shows a high maximum amplification (see also Witt and Mao (1994)).

(ii) The gravitational lensing of extended sources could induce a variable polarisation during the lensing event. This was first discussed by Schneider and Wagoner (1987) in relation to supernovae, and recently by (Simmons et al. 1994) in relation to microlensing surveys. The mechanism is the following. Light emerging from the limb of the star will normally be polarised, as was first discussed by Chandrasekhar (1960). During the microlensing of the star there will be an amplification of the light from the limb of the star nearest to the projected position of the lens, and hence the integrated light from the source should display a net polarisation. Furthermore the position angle of the polarised light would simply depend on the angle made by the line joining the source and lens centres projected onto the source plane. Chandrasekhar's calculations for an electron atmosphere gave a limb polarisation of about 10%. More accurate calculations indicate a lower degree of polarisation in the visible band for grey atmosphere, although high polarisation could arise in stellar envelopes (Collins and Buerger1974). The maximum degree of polarisation attained during a lensing event would depend on the star's radius and the distance $d_0$, or impact parameter, but would typically be less than than 10%. Successful measurement of this variable polarisation yields still more information about the lensing geometry and provides a further means of determining the mass and distance of the lensing object.

One obvious problem with this method is the degree of polarisation and brightness of the star required for the effect to be observable. It is however an interesting feature of this approach that the rise in polarisation will usually take place later than rise in total flux. Typically, since the polarisation variability is a differential effect, somewhat akin to tidal forces, the period over which the variable polarisation takes place will be half that of the flux variability. Thus it would be possible to implement an early warning system: use the detection of flux variability to alert a search for polarisation variability on a larger telescope.

(iii) Finally, although for point sources and for extended sources with no limb darkening the light curves should be independent of the wavelength of light, or achromatic, it is possible that chromatism could come about through the wavelength dependency of limb darkening. For most situations we would not expect this to be an important effect, however.

With the detection of more events, the interpretation of their statistics will provide more information about the distribution of lensing objects. With so few events confirmed a full analysis is probably premature, and would require a full understanding of the selection effects at play. In this paper we shall only make a few remarks about the fraction of events that would be expected to show polarisation variability, and leave a fuller analysis to a future publication.

Outline of paper is as follows. In section 2 we develop the theoretical formulation of the problem and introduce our notation. Our numerical results for polarisation and flux variability are presented in section 3. In section 3.4 we discuss the possible chromatism introduced by extended sources and give numerical results for the grey atmosphere, and in section 3.5 we make a few remarks about the statistics of the events.

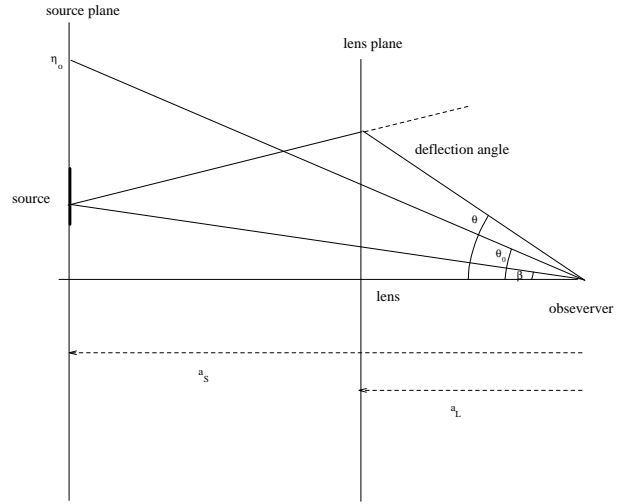

**Figure 1.** Two dimensional representation of lensing geometry.

## 2 THEORETICAL FORMULATION

In this paper we shall only consider Schwarzschild lenses, where the lensing object has a spherically symmetric gravitational field.

### 2.1 The lens equation

Two images of any point source on the source plane are formed by the lens, except in the case where the source point, lens and observer are all aligned, in which case an Einstein ring is formed. A light ray, see figure 1, is deflected through an angle $4GM/rc^2$ or $2R_S/r$ where $M$ is the mass of the lens, $R_S$ is the Schwarzschild radius of the lensing object, and $r = a_L\theta$. The apparent directions of these images are given by $\theta_1$ and $\theta_2$. It is easily shown that these satisfy the simple lens equation

$$\theta^2 - \theta\beta - \theta_0^2 = 0 \quad (1)$$

where $\beta$ is the actual direction of the source point (i.e. in the absence of lensing) and $\theta_0$ the so called Einstein radius, given by

$$\theta_0^2 = \left(\frac{1}{a_L} - \frac{1}{a_S}\right) 2R_S \quad (2)$$

where

$$R_S = mR_{S\odot} \quad (3)$$

where $m = M/M_\odot$ and $R_{S\odot}$ is the Schwarzschild radius of the Sun.

### 2.2 Amplification

We shall introduce as a natural unit of distance the Einstein radius projected onto the source plane, $\eta_0$, given by

$$\eta_0 = \xi^{-1/2}(1-\xi)^{1/2} m^{1/2} D \quad (4)$$

where $D = \sqrt{2R_{S\odot} a_S}$ and $\xi = a_L/a_S$.

This quantity $\eta_0$ is fundamental in this analysis. If the distance, $d$, of the source from the lens projected onto the source plane is much greater than $\eta_0$, there will be negligible amplification. If the actual radius of the source star is much less than $\eta_0$ then it can effectively be taken as a point source. The time duration of a lensing event can be approximately taken to be

$$2\frac{\eta_0}{\xi v_\perp} = 2\, v_\perp \xi^{-3/2}(1-\xi)^{1/2} m^{1/2} D \qquad (5)$$

where $v_\perp$ is the transverse velocity of the lens (see fig. 2).

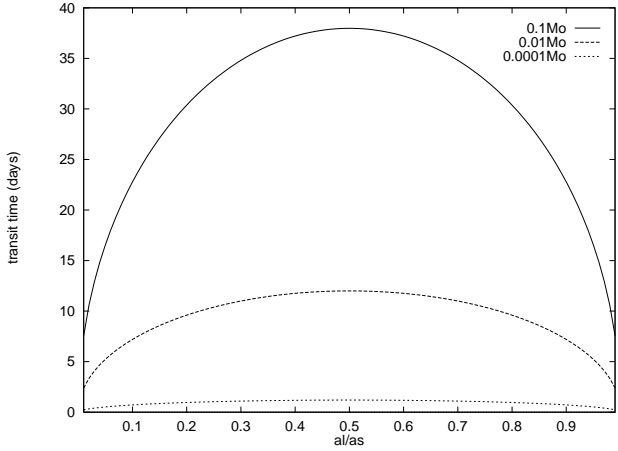

**Figure 2.** Transit times in days as a function of the fractional distance to the LMC assuming that $v_\perp = 100 kms^{-1}$ for a variety of lens masses

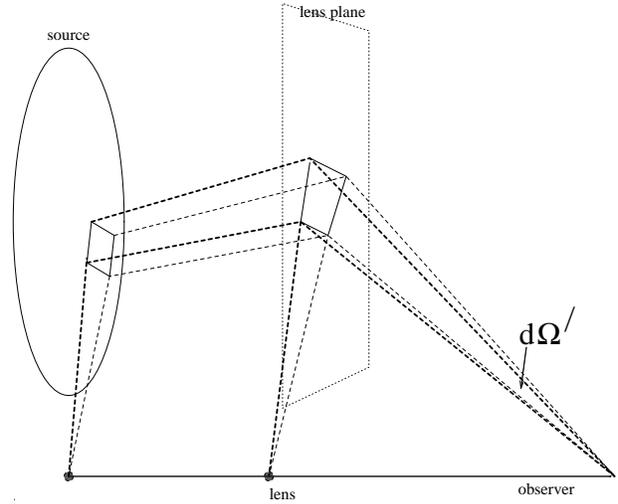

**Figure 3.** Three dimensional representation of the lensing geometry illustrating amplification

If this time is less than two days or so, there is little possibility of observing such an event. This puts constraints on the source distance and the lens mass for a given lens velocity and source distance (see figure 7).

Amplification of the source essentially arises from the fact that the apparent solid angle, $d\Omega'$ subtended at the observer by an element of area of the source is greater than the solid angle, $d\Omega$, subtended by the same element in the absence of the lens (see figure 3). The specific intensity, $I$, along the (deflected) ray, however, remains constant. Hence, with reference to fig. 3, the flux at the observer due to this element is amplified and may be expressed as

$$\begin{aligned}dF &= I(\zeta,\phi)\, d\Omega' \\ &= I(\zeta)\, \frac{d\Omega'}{d\Omega} d\Omega \\ &= I(\zeta,\phi)\, A(\zeta,\phi)\, d\Omega \qquad (6)\\ &= I(\zeta,\phi)\, A(\zeta,\phi)\, \frac{\zeta d\zeta d\phi}{a_S{}^2} \qquad (7)\end{aligned}$$

where we have of course assumed that the angles $\theta$, $\theta_0$, and $\beta$ are very much less than 1, and $A$ is the amplification. Because of the azimuthal symmetry of the lens, this amplification factor can simply be written as a function of the distance of the source element from the projected position of the lens in units of $\eta_0$, $\zeta$, and is given by $\frac{d}{d\beta}\theta$. Since for our case of microlensing the time delay between the two images is negligible, the total amplification of the source element is given by the sum of the amplification for the two images. This gives the amplification factor as

$$A(z) = \frac{1}{2}\left(z + \frac{1}{z}\right) \qquad (8)$$

where

$$z = \left(1 + \frac{4}{\zeta^2}\right)^{\frac{1}{2}} \qquad (9)$$

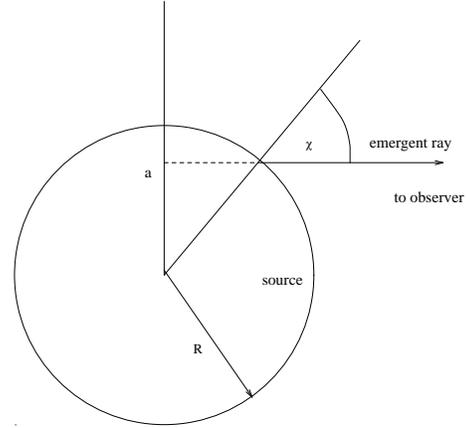

**Figure 4.** Light emerging at angle $\chi$ to the normal to the stellar surface will be limb darkened and polarised.

### 2.3 Extended source

We shall assume the star to be spherical (see figure 4). The specific intensity of the radiation that emerges from the stellar photosphere will depend on the cosine of the emergent angle, $\mu = \cos\chi$. We shall assume this to take the simple form

$$I(\mu) = i_0 + i_1\, \mu \qquad (10)$$

where $i_0$ and $i_1$ can be related to the luminosity of the star and the limb darkening (see Chandrasekhar (1960)). The flux at the stellar surface is simply

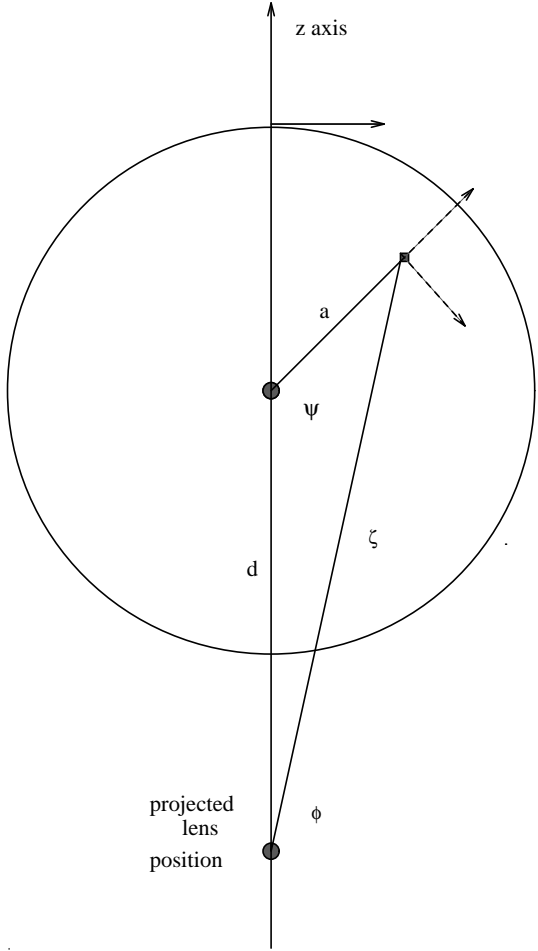

**Figure 5.** The total polarised flux is obtained by rotating the contribution from each source element through angle $\pi - \psi$ and integrating over the source

$$2\pi \int_0^1 I(\mu)\mu d\mu = \pi \left( i_0 + \frac{2}{3} i_1 \right) = \frac{L}{4\pi R^2} \tag{11}$$

where $L$ is the luminosity of the star and $R$ its radius. The ratio $i_1/i_0$ can be found by modelling stellar atmosphere using the theory of radiative transfer, and determines the degree of limb darkening. For a grey atmosphere, this ratio takes the value $\frac{3}{2}$.

The state of polarisation of this emergent radiation is best expressed in terms of the Stokes parameters $U$ and $Q$ (Chandrasekhar 1960). Light emerging at angle $\chi$ to the normal to the stellar surface will be limb darkened and polarised.

In the coordinate frame whose $z$ axis lies in the plane defined by the normal to the photosphere and the emergent direction (see figure 5) the Stokes parameters of the emergent radiation are given by

$$\begin{aligned} U &= u_0(1-\mu) \\ Q &= 0 \end{aligned} \tag{12}$$

This simply states (assuming $u_0$ to be positive) that the polarisation is in a direction perpendicular to the plane defined by the normal and the emergent direction. The degree of polarisation is given by

$$p = \frac{(U^2 + Q^2)^{1/2}}{I} \tag{13}$$

From a somewhat simplified calculation assuming Thompson scattering, Chandrasekhar obtained a value of 11.7% for the maximum polarisation, that is for light emerging at the limb.

To obtain the total flux and polarised flux at the observer we must integrate the contributions of each surface element over the source's disc. For the total flux this simply gives

$$F_I = \frac{1}{a_s^2} \int\!\!\int_{\text{disc}} (i_0 + i_1 \mu) A(\zeta) \zeta \, d\zeta d\phi \tag{14}$$

It is convenient to write equation (14) as

$$F_I = \frac{1}{a_s^2} (i_0 \alpha_0(R,d) + i_1 \alpha_1(R,d)) \tag{15}$$

where

$$\alpha_0(R,d) = \int\!\!\int_{\text{disc}} A(\zeta) \, \zeta \, d\zeta d\phi \tag{16}$$

$$\alpha_1(R,d) = \int\!\!\int_{\text{disc}} \mu \, A(\zeta) \, \zeta \, d\zeta d\phi \tag{17}$$

In order to evaluate the polarised flux at the observer, we need to evaluate $F_U$ and $F_Q$. When summing the contributions of the surface elements of the source, we must refer the stokes parameters to the same reference frame. We shall take this to have its $z$ axis aligned with the line in the source plane connecting the source's centre to the projected position of the lens (see figure 5). Using the transformation properties of the Stokes parameters we obtain in this new frame

$$\begin{aligned} dU_{\text{new}} &= dU_{\text{old}} \cos 2\psi - dQ_{\text{old}} \sin 2\psi \\ &= dU_{\text{old}} \cos 2\psi \end{aligned} \tag{18}$$

$$\begin{aligned} dQ_{\text{new}} &= dU_{\text{old}} \sin 2\psi + dQ_{\text{old}} \cos 2\psi \\ &= dU_{\text{old}} \sin 2\psi \end{aligned} \tag{19}$$

Integrating over the disc to obtain the polarised flux at the observer we obtain

$$\begin{aligned} F_U &= \frac{1}{a_s^2} \int\!\!\int_{\text{disc}} u_0(1-\mu) A(\zeta) \cos 2\psi \zeta d\zeta d\phi \\ &= \frac{1}{a_s^2} u_0(\beta_0(R,d) - \beta_1(R,d)) \end{aligned} \tag{20}$$

and

$$F_Q = \frac{1}{a_s^2} \int\!\!\int_{\text{disc}} u_0(1-\mu) A(\zeta) \sin 2\psi \zeta \, d\zeta d\phi \tag{21}$$

where

$$\mu = \left[ 1 - \left(\frac{a}{R}\right)^2 \right]^{1/2} \text{ and } a^2 = d^2 + \zeta^2 + 2\zeta d \cos\phi \tag{22}$$

and we have introduced the functions

$$\beta_0(R,d) = \int\!\!\int_{\text{disc}} \zeta \, A(\zeta) \cos 2\psi \, d\zeta d\phi$$

$$\beta_1(R,d) = \int\!\!\int_{\text{disc}} \zeta \mu \, A(\zeta) \cos 2\psi \, d\zeta d\phi \tag{23}$$

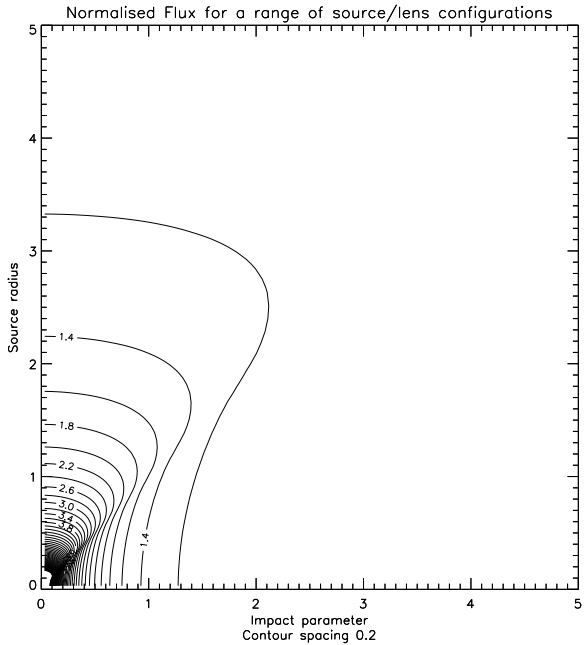

**Figure 6.** Contours in $R$ - $d_0$ plane of constant flux amplification for $i_1/i_0 = 3/2$ corresponding to the grey atmosphere. Amplification is expressed in negative magnitudes. Radius of star, $R$, and impact parameter, $d_0$, are both expressed in units of $\eta_0$.

The angle $\psi$ can similarly be expressed as a function of $\zeta$ and $\phi$. (The expression for $F_Q$ is necessarily zero). When the lens lies outside the disc, the integration is considerably simplified by taking the origin to be the centre of the disc of the source. We shall numerically evaluate these integrals and make no attempt to write down analytic forms for them, which would necessarily be very complicated.

To obtain the time profiles of the flux and degree of polarisation, we only need note that the distance, $d$, is given by

$$d = \left[ d_0^2 + \left(v_\perp \frac{a_S}{a_L} t\right)^2 \right]^{1/2} \qquad (24)$$

Here $d_0$ is the impact parameter, or distance of closest approach, and $v_\perp$ the transverse velocity of the lens.

## 3 NUMERICAL RESULTS

### 3.1 Flux as function of radius of star and distance of lens

The results of the flux integration are given in figure 6 for the cases of uniform intensity ($i_1/i_0 = 0$) and limb darkening corresponding to the grey atmosphere ($i_1/i_0 = 3/2$). The flux has been normalised to the flux in the absence of lensing. Flux contours are shown on the $R$ and $d$ plane, both expressed in units of $\eta_0$. There are a number of interesting features. For $R = 0$, the amplification is evidently that a point source. For fixed $d$ the flux increases with increases radius, until it reaches a maximum at around $R = d$. (This point was observed by Gould (1994b) and Witt and Mao (1994)). This means that the probability of detection is in fact greater than would have been supposed if we had taken the sources to be point like. However, as $R$ increases beyond $d$, the flux falls off. For values of $R > 3\eta_0$, the flux amplification is always less than 10% no matter the value of $d$. Thus such lensing events in which the source radius in greater than 3 would not be detectable according to the criteria adopted by the current surveys. This would discriminate against low mass and distant lenses (low $\eta_0$) being detected (see figure 7).

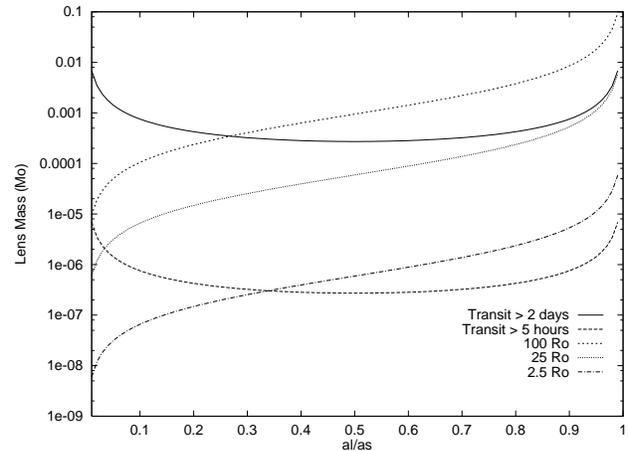

**Figure 7.** Observable lensing events are constrained by transit times and flux amplification. Graphs show the mass of lens that produces observable events as a function of fractional distance to the source (assumed to be the LMC). Only events above the curves are observable. $v_\perp$ is taken to be $100\,km\,s^{-1}$.

The time profile of the flux amplification may be obtained by considering the same figure, but now allowing $d$ to decrease to some minimum value, $d_0$, and then increase again. These time profiles will be discussed below. If $d$ attains a value less than $R$, the lens passes over the face of the star, and we shall call such an event a transit. If this does not occur, i.e. $d_0 > R$, we call the event a bypass.

### 3.2 Polarisation contours

Figure 8 shows the contours for the degree of polarisation in the $R$ and $d$ plane. Polarisation is always perpendicular to the line of centres, ie $F_U$ is positive. Once again the maximum polarisation for fixed $d$ is achieved at around $R = d$, which corresponds to the maximum amplification of the limb. Because of dilution by the rest of the source, the value will not reach the Chandrasekhar value at the limb of 11.7%. Maximum polarisation will evidently be achieved when $R$ is small but of the same size as $d$. This requires a large value of $\eta_0$. Such events would not take place very frequently. However, values of polarisation of more than 0.1% would be relatively frequent, and for bright stars (and big telescopes) could be observed.

To obtain the time profiles of polarisation, keep $R$ fixed and allow $d$ to decrease to its minimum value $d_0$ and then increase again. Bypasses will show only a single maximum, whereas transits will display a double maximum, roughly corresponding to when the lens passes over the limb. All profiles will of course be symmetric.

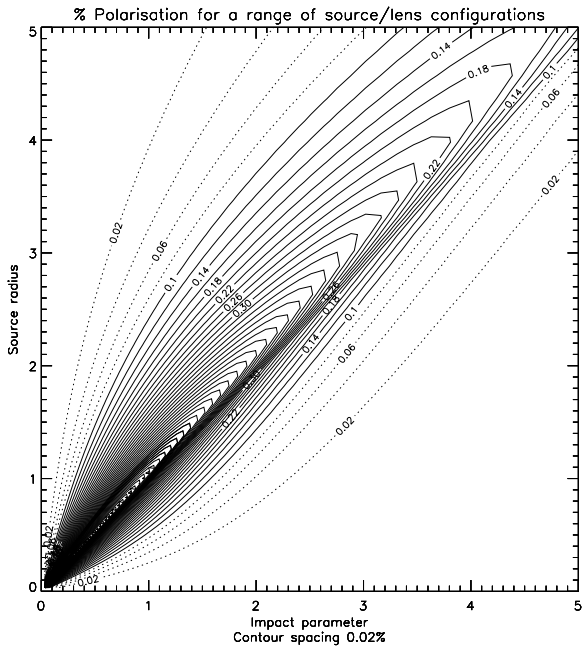

**Figure 8.** Contours of constant percentage polarisation in $R$-$d$ plane for electron scattering atmosphere with 11.7% polarisation at the limb. Radius $R$ and impact parameter $d_0$ are in units of $\eta_0$.

### 3.3 Time profiles of flux and polarisation

As we have already mentioned, lensing events of extended sources can be classed into two types, (i) transits and (ii) bypasses. In the former, the lens passes over the source whereas for bypasses it does not. Case (i) takes place when $R > d_0$ and case (ii) when $R < d_0$. The event can be recorded either as (a) variation in flux but no polarisation (b) variation in polarisation but no variation in flux (c) variation in both flux and polarisation. The profiles depend on 5 parameters for a given source distance $a_S$. These are the radius of the star $R$ and the impact parameter $d_0$ both expressed in units of $\eta_0$, and $v_\perp a_S / a_L$ where $v_\perp$ is measured in units of $\eta_0$ per second. (We could choose not to model the stellar atmosphere, but instead to determine the parameters $i_0$, $i_1$ and $u_0$ from the data).

If we assume that the radius of the star can be determined from its spectral type, then $\eta_0$ can be obtained in principle from the time variation of the flux. If we further assume a value for the actual transverse velocity of the lens, although there is no reason to do so, then the distance to the source and the mass of the lens can be determined.

The fitting of polarisation data will in addition to these parameters yield the direction of the transverse velocity. Parameter fitting will be particularly easy in the case of transits, since $d_0 \approx R \cos \Phi$ and $tv_\perp a_S/a_L \approx R \sin \Phi$ where $\Phi$ is the position angle at limb crossing.

Examples of time profiles of the flux variation and polarisation variation are given in figures 9 and 10 for various combinations of parameters, and correspond to transits and bypasses. We leave the scale for time variation arbitrary. For comparison, we show the profiles that would be obtained for a point source set-up with the same values of $d$, $M$, $\xi = a_L/a_S$ and $v_\perp$. (The flux profiles are similar to those obtained by Witt and Mao (1994)). Purely for illustrative purposes, we have taken the values of these parameters for the MACHO event quoted in the literature (Alcock et al. 1994).

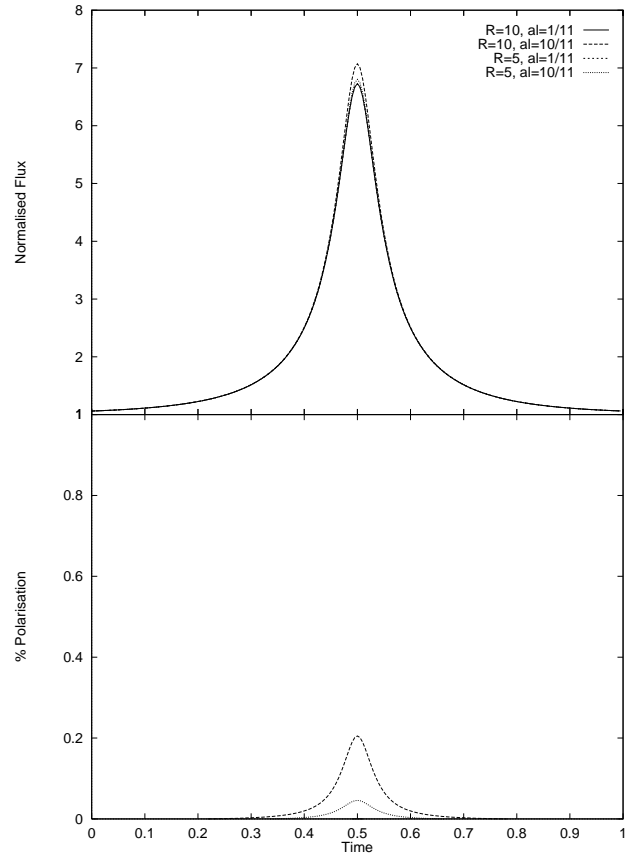

**Figure 9.** Flux and polarisation profiles for a lens of $0.1 M_\odot$ and source radii of 10 and 5 $R_\odot$ for fractional lens distance, $\xi = a_L/a_S$, of 1/11 and 10/11. Sources are taken to be in LMC, and the time unit is arbitrary.

For this event, curve fitting of flux data by Alcock et al. (1994) gave a maximum amplification of $A = 6.86 \pm 0.11$, implying (from eqn. 8) a value of $d_0 = 0.15$ (in units of $\eta_0$), and a duration time defined through eqn. (5) of $33.9 \pm 0.26$ days. These authors also inferred a most probable value of $M = 0.12\, M_\odot$ for the mass of lens. Let us take the lens mass of Alcock et al. and the value of the impact parameter for this event to illustrate the utility of polarisation and flux measurements for determining the lens geometry. In fact the time profile of the flux variation would be identical for a lens of the same mass but at a distance of $1 - a_L/a_S$, that is for a lens much closer to the LMC in this case if the source was a point source. For extended sources this will not be the case. We take $a_L/a_S$ to be 1/11 and 10/11 and the radius of the source star to be 10 and 5 $R_\odot$. $v_\perp$ is taken to be the same for all events. The corresponding profiles are shown in figures 9

The flux profiles for a nearby lens ($\xi = a_L/a_S = 1/11$) and distant lens ($\xi = 10/11$ are indistinguishable for both source radii and effectively the same as for a point source

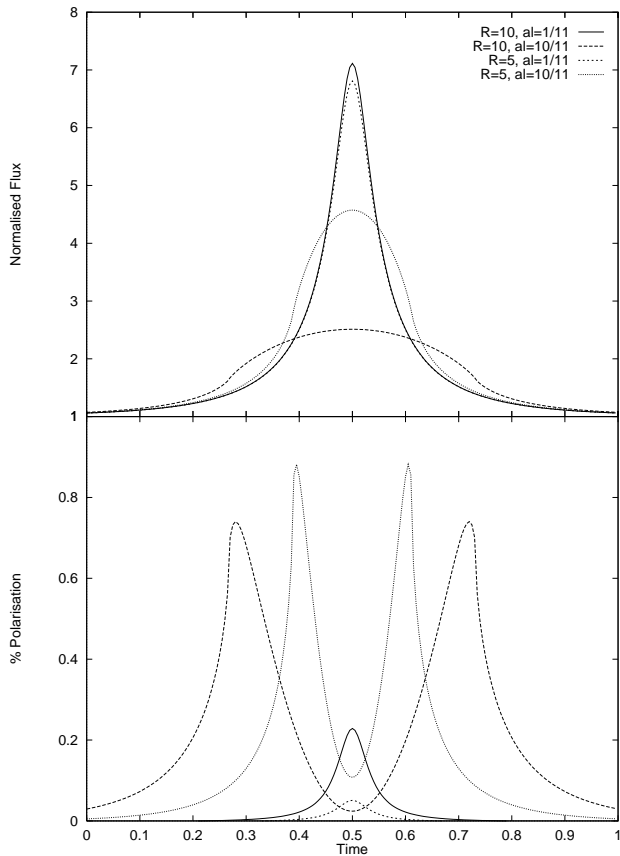

**Figure 10.** As in figure 9 but for lens of $0.001 M_\odot$. Notice that for most cases $\xi = 1/11$ and $\xi = 10/11$ are now distinguishable.

since the Einstein radius is an order of magnitude larger than the stellar radii.

The curve displaying the greatest amplification corresponds to a nearby bypass. The amplification is measurably larger than for a point source (Witt and Mao 1994). The polarisation is negligible for pointlike sources, but shows single maxima characteristic of bypasses for the two cases where the lens is near the source and rises after a measureable change in flux.

If we take the lens mass to be only $0.001 M_\odot$, the situation is quite different (see figure 10). Here the whole gamut of behaviour is displayed. Transits are now seen as flat topped flux profiles, and double maxima polarisation profiles. From the polarisation profiles for transits it is immediately apparent that one source is twice the radius of the other. For most of these cases the polarisation should be measureable.

Characteristically the rise in polarisation takes place later than the rise in the flux. This could be important in setting up an early warning system. Identification of an increase in flux could be used to signal a search for variable polarisation.

### 3.4 Chromatism

Any phenomenon in the stellar atmosphere that is limb dependent will be affected by gravitational microlensing, as

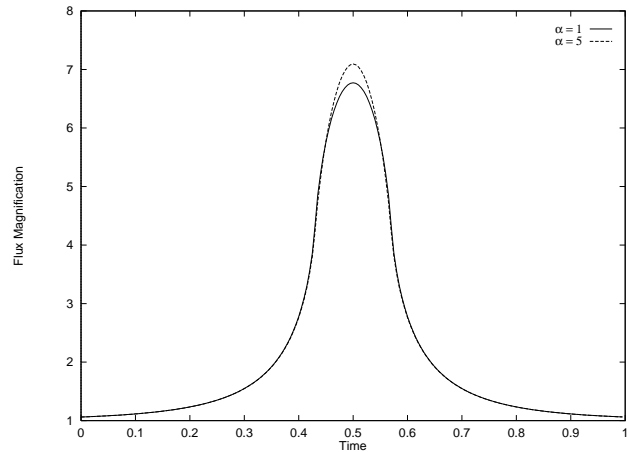

**Figure 11.** Flux amplification profiles at different wavelengths for grey atmosphere for $\alpha = h\nu/kT = 1$ and $5$.

we saw in the case of the flux and polarisation. Consider the case of the monochromatic flux, $F_\nu$, and suppose for example that the specific intensity, $I_\nu(\mu)$, is a function of the cosine, $\mu$, of the emergent angle

The flux amplification at any given frequency will then be given by

$$A_\nu(d.R) = \frac{F_{\nu_1}(d,R)}{F_{\nu_1}(\infty,R)} = \frac{\int\int_{\text{disc}} I_\nu(\mu) A(\zeta)\, \zeta\, d\zeta\, d\phi}{\int\int_{\text{disc}} I_\nu(\mu)\, \zeta\, d\zeta\, d\phi} \qquad (25)$$

If the specific intensity had the separable form $I_\nu = f(\nu)G(\mu)$ then evidently the amplification given by equation (25) would be frequency independent, and no chromatism would be observed. If on the other hand this is not the case, a degree of chromatism can be expected. How large this effect is will of course depend on the nature of the stellar photosphere. In the idealised case of a grey atmosphere, where the opacity is taken to be frequency independent and where LTE is assumed, the solution of the equation of radiative transfer

$$\mu \frac{\partial}{\partial \tau} I_\nu = I_\nu - B_\nu(T(\tau)) \qquad (26)$$

yields

$$I_\nu(0,\mu) = \int_0^\infty e^{-\tau/\mu} B_\nu(T(\tau)) \frac{d\tau}{\mu} \qquad (27)$$

where $B_\nu(T(\tau))$ is the Planck function and $T(\tau) = (1/2)^{1/4} T_e (1 + \frac{3}{2})^{1/4}$. $T_e$ is the effective temperature of the star.

The chromatism is not strong, and is only evident for small values of $d$ (see figure 11). (However, both the assumptions that the opacity is frequency independent and that LTE obtains are dubious for realistic stars). Thus it is unlikely that this source of chromatism would affect the selection of lensing events. It might possibly be used as an additional technique for determining the lensing parameters.

### 3.5 Statistics of events

A simple expression can be obtained for the rate at which events occur (e.g. Paczynski (1986)). The number of such events that are transits or bypasses, polarised/unpolarised for a fixed radius of the source is proportional to the length of the corresponding interval in the values of $d_0$, or impact parameter, at which the different types of events can take place. If we assume that all events where the amplification of the flux is greater than 1.34 are detected, and similarly all events showing a maximum polarisation of more than 0.1% are recorded as polarisation events, then the fraction of recorded events of different types can be obtained from figures 6 and 8. If for instance, the source radius is 100 $R_\odot$, and the MACHOs are assumed all to have an Einstein radius of 500 $R_\odot$, then $\sim$ 40% are polarised. For $R > 3\eta_0$, there will only be polarisation events, whereas for $R << \eta_0$, only a small fraction of events will be polarised. The distribution of events of different types will thus yield information about the spatial and mass distribution of the lensing objects, and the size distribution of the stellar sources.

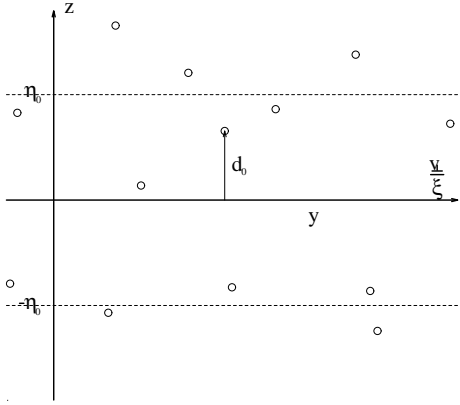

**Figure 12.** Any stellar source will provide an event during time interval $(0, T)$ if it lies within the strip $0 < y < v_\perp T/\xi$ and satisfies selection criterion.

In fact the predicted rate of recorded flux events for extended sources will differ considerably from the predicted rate for point sources. No flux event will be recorded if the radius of the star is greater than about three times the Einstein radius, $\eta_0$. This effectively acts as a selection on the mass of the gravitational lenses that will be observed.

Consider the case of a gravitational lens at distance $a_L = \xi a_S$ of mass $m$. We shall also take the radius of the source stars to be fixed. In a time interval, $T$, assumed long compared with the typical lens transit time, the number of recorded events will be simply given by

$$\int_0^{T \frac{a_S}{a_L} v_\perp} dy \int_{-\infty}^{\infty} dz \, \Sigma(y, z) \, S(\eta_0, R, z, v_\perp) \quad (28)$$

where $\Sigma(y, z)$ is the surface number density of source stars (and $|z| = \eta_0 d_0$).

$S(\eta_0, R, d_0, v_\perp)$ is the selection function. If we assume for simplicity that selection is only on magnification amplification, then for a point source $S = \Theta(\eta_0 - d_0)$. For an extended source this is no longer the case. Flux amplification will surpass 1.34 roughly when $R < \alpha \eta_0$ and $\eta_0 - d_0 < 0$. ($\alpha \approx 3$) (see figure (12)). Thus in this case the selection function may be written $S \approx \Theta(\eta_0 - d_0)\Theta(\eta_0 \alpha - R)$, where $\Theta$ is the Heaviside function. In the case where we assume the surface density is homogeneous, we may write the above integral 28 as

$$\frac{T}{\xi} v_\perp \Sigma_{S0} \int_{-\infty}^{\infty} dz \, S(\eta_0, R, z, v_\perp) \quad (29)$$

For a point source this gives an event rate of

$$\frac{v_\perp}{\xi} \Sigma_{S0} \int_{-\infty}^{\infty} dz \, \Theta(\eta_0 - |z|) = \frac{2}{\xi} v_\perp \Sigma_{S0} \eta_0 \quad (30)$$

and for an extended source, with the approximate selection function given above, yields an event rate of

$$2 \frac{v_\perp}{\xi} \Sigma_{S0} \eta_0 \Theta(\alpha \eta_0 - R)$$

If we now assume that the number density of lenses, $n_L$, depends only on the radial distance to the sources (LMC say) then we may write the total rate of events as

$$2\langle v_\perp \rangle a_S \int_{\Omega_{LMC}} a_S^2 n_{S0} \, d\Omega \int_0^1 d\xi \, \xi \, n_L(\xi) \, \eta_0(\xi) \, \Theta(\alpha \eta_0(\xi) - R)$$

$$= 2\langle v_\perp \rangle a_S N_S \int_0^1 d\xi \, \xi \, \eta_0(\xi) \, n_L(\xi) \, \Theta(\alpha \eta_0(\xi) - R) \quad (31)$$

where $\langle v_\perp \rangle$ is the mean transverse velocity. Now $\eta_0$ is given by equation (4) as a function of $m$ and $\xi$. If we take the number density of lenses to be constant, then this expression reduces to

$$2\langle v_\perp \rangle a_S N_S n_L \int_0^{\frac{1}{1+\rho^2}} d\xi \, \xi \, \eta_0(\xi) \quad (32)$$

where $\rho = R/(\alpha m D)$ is a dimensionless radius of the star and $D = \sqrt{2 R_{S\odot} a_S}$ as before. This can be integrated analytically to yield

$$2\langle v_\perp \rangle a_S N_S n_L D F(\rho) \quad (33)$$

where

$$F(\rho) = \frac{1}{4} \left\{ \cot^{-1} \rho - \frac{\rho(\rho^2 - 1)}{(1 + \rho^2)^2} \right\}$$

A normalised plot of this function is given in figure (13), and shows the fraction of lensing events that would actually be detected if the star was an extended source compared with the number predicted assuming that the source was point-like. For source stars of 10 solar radii and $\alpha = 3$ only about 5% of events corresponding to lens masses of $10^{-4}$ solar masses will be picked up, whereas for lens masses of 0.1 solar mass about 99% will be detected.

This analysis could be trivially extended to cover the case where there was a distribution of lens masses, and a distribution of stellar radii of source stars.

(Obviously, the rate at which events are detected will depend on a number of selection effects, which we shall not attempt to model).

## 4 CONCLUSIONS

Microlensing of stellar sources should produce variable polarisation. Depending on the geometry of the lensing system, this could be as high as 2%. A significant fraction of MACHO events in which stars from the LMC are lensed and manifesting flux variation should also display variable

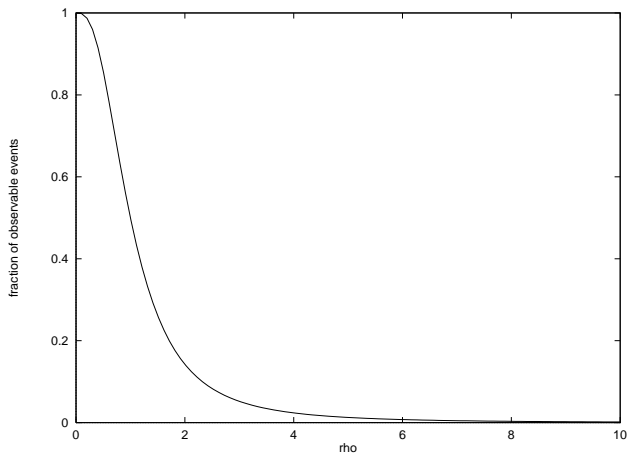

**Figure 13.** If the sources are extended rather than pointlike then fewer lensing events will take place. The fraction of observable events for an extended source compared with the predicted number of observable events for pointlike source as a function of the dimensionless source radius, $\rho = R/(\alpha m D)$

polarisation above 0.1% polarisation, although the precise number will depend on the size distribution of stars, and the mass and spatial distribution of lenses. Evidently, for nearby lenses the fraction will be small, and polarisation events are unlikely. For lensing objects lying closer to the LMC, polarisation events and transits will be more frequent.

In principle the measurement of the variable polarisation yields the Einstein radius of the lens and its velocity direction on the sky, and thus provides more information than simple measurement of the flux variation assuming an extended source. Transits would be easily distinguished by their double peak polarisation profile.

The low levels of polarisation produced by lensing mean that the stars would have to be sufficiently bright for a given size of telescope and given integration time. It should be feasible to measure 0.1% polarisation of a star brighter than $18^{th}$ magnitude with one hour's integration time on a 1 metre telescope. The intrinsically brighter stars are more likely to display higher levels of limb polarisation and are also more amenable to observation.

During lensing the rise in polarisation generally takes place significantly later than the rise in flux. This provides the opportunity of implementing an early warning system for detailed study of the polarisation variability. This also in part offsets the difficulty in observing low degrees of polarisation for faint stars, since a larger telescope could be brought in for the observation of polarisation. Since we are here dealing with variable polarisation, the effect of interstellar polarisation could also be accounted for.

For high mass lenses variable polarisation would be more likely to be observed for more extensive stars. There is also the interesting possibility of probing with gravitational lenses the stellar atmosphere and envelopes, which can be quite extensive, of bright young stars.

We have presented the simplest model for limb polarisation in stars. More detailed modelling of the stellar photosphere for stars of given spectral type should allow more accurate determination of the lensing parameters, although in principle the parameters determining the limb dependency of the polarisation could be fitted as well. Thus polarisation measurement should be a valuable supplementary tool in the analysis of microlensing by compact objects in the galaxy.

## ACKNOWLEDGEMENTS

AMN acknowledges the support of a PPARC studentship and JPW the support of a Cormack Vacation Scholarship Research Scholarship. The authors would like to thank Andy Gould, Andrew Conway and George Collins for useful comments.

## REFERENCES

Alcock, C., Akerlof, C.W., Allsman, R.A., Axelrod, T.S., Bennett, D.P., Chan, S., Cook, K.H., Freeman, K.C, Griest, K., Marshall, S.L., Park, H.-S., Perlmutter, S., Peterson, B.A., Pratt, M.R., Quinn, P.J., Rodgers, A.W., Stubbs, C.W., and Sutherland, W., 1994, Nature 365, 621
Chandrasekhar, S., 1960, Radiative Transfer, Dover
Collins, G.W. II and Buerger, P.F., 1974 in Gehrels, T., ed, Planets, Stars and Nebulae. Univ. Arizona Press, Tucson, p. 663
Gould, A., 1994a, ApJ 421, L75
Gould, A., 1994b, ApJ 421, L71
Maoz, D. and Gould, A., 1994, ApJ 425, L67
Nemiroff, R.J. and Wickramasinghe, W., 1994, ApJ 424, L21
Paczynski, B.: 1986, ApJ 304, 1
Schneider, P. and Wagoner, R., 1987, ApJ 314, 154
Simmons, J.F.L., Willis, J.P., and Newsam, A.M., 1994, A&A Letters, In press
Witt, H.J. and Mao, S., 1994, ApJ 430, 505